\documentstyle[twoside,fleqn,espcrc2,epsfig]{article}

\newcommand{\AmS}{{\protect\the\textfont2
  A\kern-.1667em\lower.5ex\hbox{M}\kern-.125emS}}
\def\frac#1#2{ {{#1} \over {#2} }}

\def\gtap{\raisebox{-.4ex}{\rlap{$\sim$}} \raisebox{.4ex}{$>$}}
\def\VEV#1{\left\langle #1\right\rangle}

\def\ie{\hbox{\rm i.e. }}

\def\beq{\begin{equation}}
\def\eeq{\end{equation}}
\newcommand{\ba}         {\begin{eqnarray}}
\newcommand{\ea}         {\end  {eqnarray}}
\newcommand{\ban}        {\begin{eqnarray*}}
\newcommand{\ean}        {\end  {eqnarray*}}
\def\re#1{(\ref{#1})}
\def\L{\Lambda}

\def\b{\beta}
\def\r{\rho}

\def\g{\gamma}
\def\as{\alpha_{\sf s}}

\def\bl{\beta_{\rm lat}}
\def\cl{c^{\rm lat}}
\def\cpr{c^{\rm ren}}

\def\Wp{W^{\rm pert}}

\def\Tr{\mbox{Tr}\;}

\def\np#1#2#3{Nucl.\ Phys.\ B#1 (19#3) #2}
\def\pl#1#2#3{Phys.\ Lett.\ #1B (19#3) #2}
\def\pr#1#2#3{Phys.\ Rev.\ D #1 (19#3) #2}
\def\prep#1#2#3{Phys.\ Rep.\ #1 (19#3) #2}
\def\prl#1#2#3{Phys.\ Rev.\ Lett.\ #1 (19#3) #2}

\title{Power corrections and perturbative coupling from lattice gauge
theories}

\author{G. Burgio\address{Dipartimento di Fisica, Universit\`a di Parma 
	and INFN, Gruppo Collegato di Parma, Italy}\thanks{Speaker at the 
		conference.}, 
	F. Di Renzo\address{Department of Mathematical Sciences,
	University of Liverpool, United Kingdom}, 
	G. Marchesini\address{Dipartimento di Fisica, Universit\`a di Milano
	and INFN, Sezione di Milano, Italy}       
        and E. Onofri$^{\rm a}$}

\begin{document}

\begin{abstract}
From the analysis of the perturbative expansion of the lattice 
regularized gluon condensate, toghether with MC data, we present
evidence of OPE-unexpected dim-2 power corrections in the scaling 
behaviour of the Wilson loop. These can be interpreted as an indication that 
in lattice gauge theories the running coupling at large momentum contains
contributions of order $\Lambda^2/Q^2$. 
\end{abstract}

\maketitle

\section{The gluon condensate}

Operator product expansion (OPE) \cite{W} has been applied
to the study of non-perturbative contributions of physical
observables in asymptotically free theories \cite{ITEP}. In these studies an
important r\^ole is played by the gluon condensate $\VEV {\as
\Tr\;F^2}$, which, according to OPE, has the expansion
\beq\label{OPE1} W \;\equiv\; \frac{\VEV{\as\;\Tr F^2}}{Q^4} \;=\; W_0
+ ({\L^4}/{Q^4})\;W_4 + \cdots \,, \eeq where $\L$ is the physical
scale of the theory related to the running coupling $\as=\as(Q^2)$ at
the scale $Q$.  Since $\L$ has no expansion in $\as$, perturbative
contributions are present only in $W_0$, which from power counting is
quartically divergent in the UV region.  The term with the power
$\L^2/Q^2$ is absent since there are no gauge invariant operators of
dimension two.

In the lattice theory all frequencies are bounded by the UV cutoff
$Q=\pi/a$ with $a$ the lattice spacing.  The condensate $W$ can be
written in the general form (we assume an infinite lattice for the
moment) 
\beq\label{Int1} W =\int^{Q^2}_0\; \frac{k^2\,dk^2}{Q^4}
\;f(k^2/\L^2) \,. 
\eeq 
This expression is based on the fact that the associated observable
has dimension four and is renormalization group invariant \cite{ITEP1}
so that the function $f(k^2/\L^2)$, which does not depend on $Q$, for
large $Q$, can be expressed in terms of a running coupling at the
scale $k^2$.

In \cite{ours} it is shown that the ``perturbative'' contribution, obtained
using 
for $f$ the two loop $\as(k^2)$ and introducing a cut-off $\r\L^2$
with $\r \gg 1$ to stay within the perturbative region, 
can be expressed as 
\ba
W_0^{\rm ren} & =& {\cal N} \int_{0}^{z_{0_-}} dz\;e^{-\b z}
\;(z_0-z)^{-1-\g}\nonumber\\
&=& \sum_{\ell=1}\;\b^{-\ell}\;\{\cpr_\ell
\;+\;{\cal O}( e^{-z_0\b} )\}\label{Wren}
\ea
where the integration region is mapped into
$0<z<z_{0_-} = z_0\,(1-\bar\b/\b)$ with 
$6/\bar \b = 4\pi \as(\r\L^2)$,
$e^{-z_0\b} \sim \L^4/Q^4$ and $\cpr_\ell$ the renormalon 
coefficients
\beq\label{cpr}
\cpr_\ell ={\cal N'}\; \Gamma(\ell+\g)\;z_0^{-\ell}
\,.
\eeq
A similar factorial growth of the perturbative coefficients is found
if one considers the contributions from higher powers of the coupling.
Then the expression above gives a general form of the perturbative
factorial growth with the numerical constant ${\cal N}$ which takes
into account higher order corrections.

The coefficients $\cl_\ell(M)$ of the perturbative expansion of $W(M)$ 
on a lattice of size $M$
\beq\label{pert}
\Wp(M) \;=\; \sum_{\ell\ge1} \; \cl_\ell(M) \; \bl^{-\ell}
\,,
\eeq
are known up to eight loops. The first three terms have been computed 
analytically \cite{Pisa} for an infinite lattice.
Eight terms for the expansion of both $W_{1\times1}$ and
$W_{2\times2}$ have been computed numerically in Ref.~\cite{DMO}
for a lattice with $M=8$.

In Ref.~\cite{DMO} it has been shown that the growth with $\ell$ of
the first eight coefficients is consistent with the factorial behaviour
described in the previous section for a quantity of dimension
four like $W$. The analysis of finite volume effects and the comparison
between continuum and lattice schemes have been performed in \cite{ours},
the former by taking into account an explicit IR cutoff $Q_0=2\pi/Ma$ 
\ie $z<z_{\rm ir} =4\ln (M/2) /\b $ in \re{Wren}, the latter, in analogy
with the result for $\sigma$-models \cite{DMO1}, by assuming that the
scale entering in the running coupling is $s k^2 < k^2$, with $s$ such to 
rescale the $\L$ parameter from the continuum to lattice scheme. 

\section{Evidence of a $\L^2/Q^2$ contribution}

Here we present the evidence that in lattice gauge theory the
condensate contains terms of order $\L^2/Q^2 \sim e^{-z_0\b/2}$. We
analyze $W-W_0$ as a function of $\b$. The contribution $W$ is
obtained by the Monte Carlo simulation on a $8^4$ lattice \cite{MC}. 
The contribution $W_0$ is
constructed by adding to the computed eight-loop perturbative terms the rest
of the perturbative expansion,
since for large orders the perturbative coefficients of
$W_0^{\rm ren}(M)$ approach the lattice ones \cite{ours,DMO}. 

In Fig.~1a we plot the quantity
\beq\label{Wsub}
\Delta_L W(M)\;=\; W(M) \;-\; \sum_{\ell=1}^{L}\cl_{\ell}(M)\;\bl^{-\ell}
\,,
\eeq
in the range $\bl=6-7$ for various values of $L \le 8$.

\begin{figure}[htb]
\vspace{9pt}
\begin{center}
\mbox{{\epsfig{figure=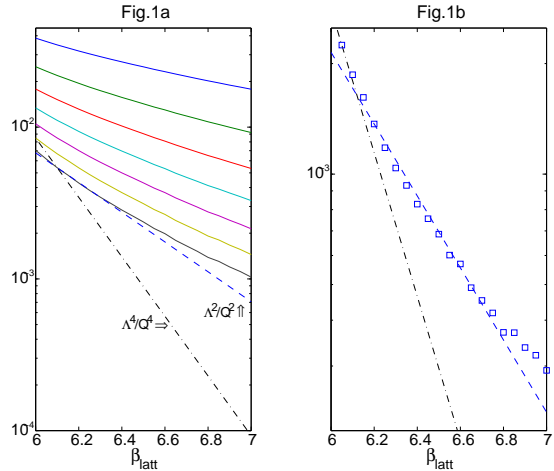,width=7.5 cm}}}
\end{center}
\caption{(a) Coefficients of the perturbative expansion 
as function of the (loop) order $\ell$ compared with the dim=2 and dim=4 
scaling. (b) Subtraction of the whole resummed perturbative expansion.}
\label{fig1}
\end{figure}

We observe that in this range of $\bl$ the quantity $\Delta_L W(M)$ approaches 
for $L\to8$ the behaviour of $\L^2/Q^2$ instead than the expected behaviour 
of $\L^4/Q^4$.

In Fig.~1b it is shown the effect of the subtraction of the perturbative 
expansion to all orders \cite{ours}. The behaviour $\L^2/Q^2$ is mantained.

As for finite volume effects, while those on the perturbative coefficients 
are under control \cite{ours}, those on the MC data are quite difficult to 
estimate without 
performing a direct 
Monte Carlo simulation on lattices with $M$ sufficient large to have 
the IR cutoff below the Landau singularity, \ie\  $\ln (M/2) > \b/12b_0$.

\section{Conclusion}

We have studied the contribution $W_0$
obtained from the first eight terms of the perturbative 
expansion and a remainder constructed on the hypothesis that 
only $\L^4/Q^4$ corrections are present.
By subtracting from $W$ the term $W_0$ we have found indications of 
an additional contribution proportional to $\L^2/Q^2$ (Fig.~1).

The reason for the unexpected behaviour $\L^2/Q^2$ could be
that the analysis is not complete. The major problem is the finiteness
of the lattice size. The effects on the
considered contributions to $W_0$ have been extimated and they seem to be 
small \cite{ours}.
It may be that the Monte Carlo simulation for $W$ contains
spurious finite size effects giving an effective $\L^2/Q^2$ behaviour.
Excluding this possibility would require an investigation on
a very large lattice with $\ln(M/2)\;\gtap\;\b$.

In lattice gauge theory,
due to the presence of a rigid UV cutoff, the gluon condensate has no 
UV renormalon \cite{Ren}. One may worry than that OPE could be violated.
Recently it has been argued by Grunberg \cite{Gr} and by
Akhoury and Zakharov \cite{Zak} that terms of order $\L^2/Q^2$
can be present in the gluon condensate which are not accounted for
by OPE, but are due to power corrections in the running coupling at
high momentum. 
 In physical schemes 
\cite{DMW,BLM} highly subleading power corrections
at large momentum are naturally present in the running coupling.
These corrections could be responsible for the appearance
of $\L^2/Q^2$ terms in the condensate due to the fact that
the integral for $W$ is quartically divergent. 
A $\L^2/k^2$ contribution in $\as(k^2)$ in the integral \re{Int1} 
gives two terms. The first of order $\L^2/Q^2$ comes from the
UV region ($k^2\approx Q^2$), the second, of the canonical 
order $\L^4/Q^4$, comes from the IR region ($k^2\approx
\r \L^2$) and mixes with the term predicted by OPE. In Fig.~2 we show a fit 
of the subtracted data with both 
contributions at the same time.
 \begin{figure}[htb]
\vspace{9pt}
\begin{center}
\mbox{{\epsfig{figure=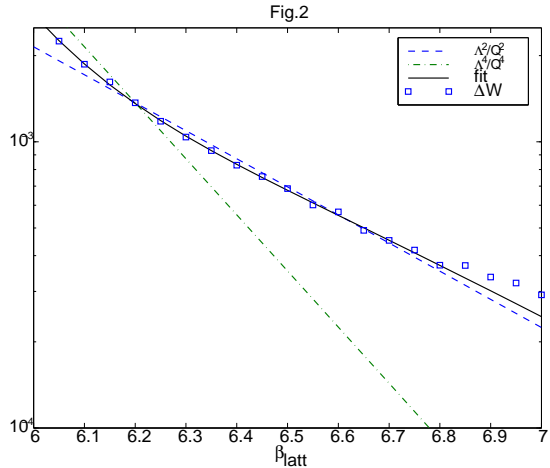,width=7.5cm}}}
\end{center}
\caption{The subtracted data fitted with a behaviour of the type $ a\, 
\L^2/Q^2+b\, \L^4/Q^4$}
\end{figure}
The $\L^2/Q^2$ terms are of ``perturbative'' nature
and are then naturally associated to the contribution $W_0$
in the OPE. Moreover they should be process independent as the running
coupling. An important question is whether these $\L^2/Q^2$ are
phenomenologically relevant (see \cite{Gr,Zak}).



\begin{thebibliography}{99}
\bibitem{W}
          K. Wilson, \pr{179}{1499}{69}.
\bibitem{ITEP}
          M.A. Shifman, A.I. Vainstein and V.I. Zakharov,
          \np{147}{385, 448, 519}{79};         
          V.A. Novikov, M.A. Shifman, A.I. Vainstein and V.I. Zakharov,
          \prep{116}{105}{84};
\bibitem{ITEP1}
         V.A. Novikov, M.A. Shifman, A.I. Vainstein and V.I. Zakharov,
         \np{249}{445}{85}.
\bibitem{ours}G. Burgio, F. Di Renzo, G. Marchesini and E. Onofri 
	hep-ph/9706209, submitted to Nucl. Phys. B.
\bibitem{Pisa}
        B.\ All\'es, M.\ Campostrini, A.\ Feo and
        H.\ Panagopoulos, \pl{324}{443}{94}
        and references therein.
\bibitem{DMO}
        F. Di Renzo, G. Marchesini and E. Onofri,
        \np{457}{202}{95}.
\bibitem{DMO1}
        F. Di Renzo, G. Marchesini and E. Onofri,
	\np{497}{435}{97}.
\bibitem{MC}
	MC data from L. Scorzato (1997).
\bibitem{Ren}
         For reviews and classic references see:\\
         V.I. Zakharov, \np{385}{452}{92};\\
         A.H. Mueller, in {\it QCD 20 years later}, vol.~1 
         (World Scientific, Singapore 1993).
\bibitem{Gr}
        G.Grunberg, hep-ph/9705290.
\bibitem{Zak}
        R. Akhoury and V.I. Zakharov, hep-ph/9705318;
        A.I. Vainstein and V.I. Zakharov, \prl{73}{1207}{94}; 
        \pr{D54}{4039}{96};
        K.K. Yamawaki and  V.I. Zakharov, hep-ph/9406373.
\bibitem{DMW}
         Yu.L. Dokshitzer, G. Marchesini and B.R. Webber,
         \np{469}{93}{96}
\bibitem{BLM} S.J. Brodsky, G.P. Lepage and P.B. Mackenzie,
\pr{28} {228} {83}.
\end{thebibliography}
\end{document}